# Efficiently Training Vision Transformers on Structural MRI Scans for Alzheimer's Disease Detection

Nikhil J. Dhinagar*, *Member, IEEE*, Sophia I. Thomopoulos, Emily Laltoo and Paul M. Thompson

*Abstract*— Neuroimaging of large populations is valuable to identify factors that promote or resist brain disease, and to assist diagnosis, subtyping, and prognosis. Data-driven models such as convolutional neural networks (CNNs) have increasingly been applied to brain images to perform diagnostic and prognostic tasks by learning robust features. Vision transformers (ViT) - a new class of deep learning architectures - have emerged in recent years as an alternative to CNNs for several computer vision applications. Here we tested variants of the ViT architecture for a range of desired neuroimaging downstream tasks based on difficulty, in this case for sex and Alzheimer's disease (AD) classification based on 3D brain MRI. In our experiments, two vision transformer architecture variants achieved an AUC of 0.987 for sex and 0.892 for AD classification, respectively. We independently evaluated our models on data from two benchmark AD datasets. We achieved a performance boost of 5% and 9-10% upon fine-tuning vision transformer models pre-trained on synthetic (generated by a latent diffusion model) and real MRI scans, respectively. Our main contributions include testing the effects of different ViT training strategies including pre-training, data augmentation and learning rate warm-ups followed by annealing, as pertaining to the neuroimaging domain. These techniques are essential for training ViT-like models for neuroimaging applications where training data is usually limited. We also analyzed the effect of the amount of training data utilized on the test-time performance of the ViT via data-model scaling curves.

*Clinical Relevance*— The models evaluated in this work could be trained on neuroimaging data to assist in diagnosis, subtyping and prognosis of Alzheimer's disease.

## I. INTRODUCTION

According to the World Health Organization (WHO) [1], around 55 million people have dementia and this number is expected to rise to 78 million by 2030; 60-70% of these are patients with Alzheimer's disease (AD) (World Health Organization 2022). AD is a progressive degenerative disease, where abnormal build-up of amyloid and tau proteins in the brain lead to progressive cell death, causing decline in memory functions and cognitive impairment. Neuroimaging techniques, including brain MRI, along with amyloid- and tau-sensitive PET, are now widely used in research to quantify AD progression and study factors that influence it. Even so, there is no single standardized neuroimaging test for Alzheimer's disease, and tools to facilitate diagnosis would be highly valuable clinically, and in research to study factors that resist or promote the disease. Deep learning methods, including convolutional neural networks (CNNs) show promise for AD diagnosis [2] [3] and related tasks such as prediction of amyloid levels [4] in the brain. Even so, CNN-based diagnostic algorithms still require large training datasets to achieve good performance, leading to interest in adapting deep learning architectures emerging in the computer vision field to medical problems.

ViTs have shown great promise in computer vision, with recent efforts to adapt them to medical imaging [5]. One key difference in the ViT architecture, compared to CNNs, is their self-attention mechanism which can capture long-range spatial dependencies, providing a more global perspective [6]. This property is intuitive since anatomical context and spatial patterns are often crucial in analyzing medical images. ViTs typically require large datasets for training, but in other contexts, this issue has been overcome by effective pre-training/fine-tuning techniques [7]. ViT and its variants have shown initial success for various neuroimaging tasks [8] [9], but typically these works train the models on thousands of MRI scans.

The main contributions of this work are that we: 1. Trained and tested models representative of the ViT class of architectures in limited neuroimaging data settings for tasks varying in difficulty i.e., sex (easier) and AD classification (more difficult), 2. Evaluated the benefit of pre-training ViTs on 100,000 synthetic 3D brain MRI scans generated by a recently proposed latent diffusion model [10] (LDM; this training dataset is much larger than any commonly used neuroimaging datasets). 3. Tested the effect of advanced training strategies such as data augmentation and learning rate warm-ups followed by annealing.

The pre-training of deep learning methods on synthetic data from a generative model is a recent innovation, and

N. J. Dhinagar, S. I. Thomopoulos, E. Laltoo and P. M. Thompson are with the Imaging Genetics Center, Mark and Mary Stevens Neuroimaging and Informatics Institute, Keck School of Medicine, University of Southern California, Los Angeles, CA, USA (*corresponding author: Nikhil J. Dhinagar).

one could generate an almost unlimited, diverse set of synthetic training data conditioned on clinical variables and benefit from improved performance via pretraining of large vision models.

## II. DATA

To train and test our algorithms, we used T1-weighted 3D brain MRI scans from the UK Biobank (UKBB) [11], the Alzheimer's Disease Neuroimaging Initiative (ADNI) [12], and synthetic scans obtained from an LDM [10]. First, for pre-training, we used 38,710 scans (age: 44.6-82.8 years, 20,219F/18,491M) from UKBB and 100,000 (44-82 years, 50,123 F/49,877 M) synthetic scans from the LDM for sex classification using our models, as advocated by Lu et al., [3]. The data was split into training, validation and test sets of size 37,176, 759 and 775 from UKBB and 98,981, 498, and 500 from the LDM. We used 4098 ADNI scans from 1188 subjects (55.7-92.8 years, 418F/461M) for AD classification; these were split into sets of size 2577, 302, and 1,219 for training, validation and testing, stratified to restrict unique subjects to specific folds. We used data from the Open Access Series of Imaging Studies [13], phase 3 (OASIS3), as the out-of-distribution dataset to independently evaluate our models. We used 600 scans (age: 43.5-97.0 years, 341F/359M) from OASIS3.

MRI data from UKBB and ADNI were pre-processed via a standard sequence of steps [2] including 'skull stripping' for brain extraction, N4 bias field correction, linear registration (with 9 degrees of freedom) to a UK Biobank minimum deformation template, and isometric voxel resampling to 2 mm. The LDM data was generated using a diffusion model that was originally trained on UKBB data [10].

The T1-weighted (T1-w) input scans were resized to 80x80x80 to ensure straightforward correspondence with the patch sizes used in the ViT models.

## III. METHODS

### A. Model Architectures

As a baseline model, we used a random forest classifier trained with radiomics features extracted from the T1-w MRI scans [2]. The radiomics features extract image descriptors based on the gray-level co-occurrence matrices using the PyRadiomics package [14].

We trained two variations of the ViT, i.e., the ViT-B/16 [15] and the neuroimage transformer, NiT [16] as illustrated in Fig.1.

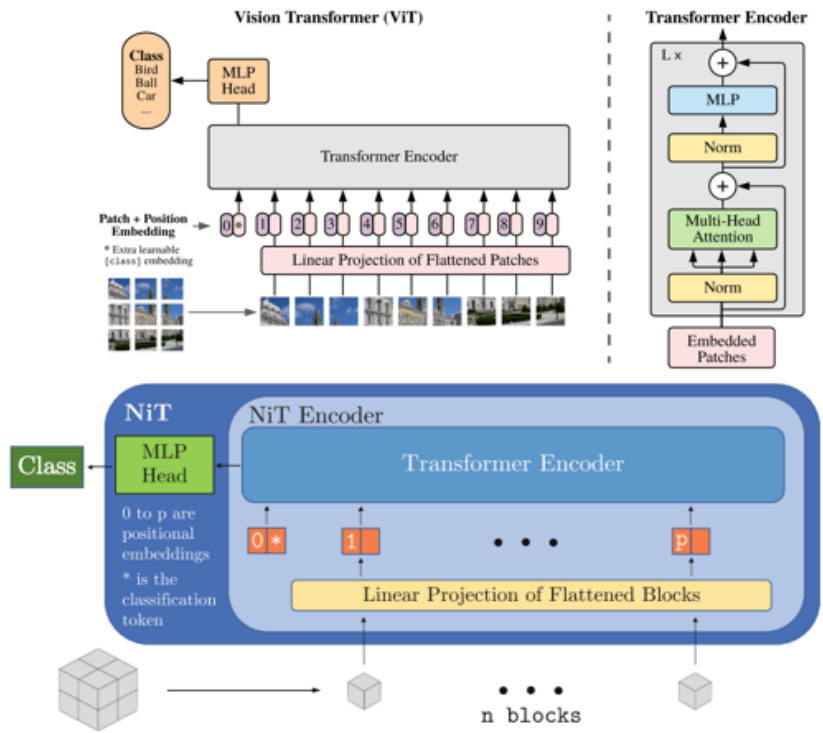

Figure 1. Overview of the Vision Transformer architecture (reproduced from [15] [16]). The input image is split into fixed-sized patch embeddings and combined along with learnable position embeddings and a class token. The resultant sequence of vectors is fed into a transformer encoder, which consists of alternating layers of multi-head attention and an MLP, as shown on the top right. The ViT-B/16 architecture uses a patch size of 16, but the original architecture was modified to create patches from 3D scans (instead of just 2D images). The NiT architecture is a scaled-down version of the ViT-B/16, with fewer encoder layers and attention heads.

The ViT-B/16 used a patch size of 16, hidden dimension size of 768, 12 transformer encoder layers and 12 self-attention heads with a dropout of 0.1. The NiT used a patch size of 8, hidden dimension size of 256, 6 transformer encoder layers and 2 up to 12 self-attention heads with a dropout of 0.3. The NiT architecture served as the primary model for our experiments. We optimized the default values of the NiT's number of transformer encoder layers and attention heads. Both models were used with the architectural design parameters noted but staying consistent with the originally proposed architectures to serve as reliable/reproducible benchmarks. Given the reduced numbers of layers and attention heads, the NiT model served as a scaled-down version of the ViT-B/16 to a range of 1.7 to 5M parameters (depending on the number of encoder layers and attention heads), versus the original 88M parameters.

### B. Pre-training Techniques

We utilized supervised learning via sex classification as our pre-training task as advocated by Lu et al. [3]. Prior works have shown improved classification performance for other medical imaging tasks when models are also trained on synthetically generated data [17] [18]. Here, we used both real MRI scans from the UKBB dataset, as well as synthetic scans generated by Pinaya et. al., [10] with an LDM to pre-train our ViTs. The UKBB sample size was 58 times larger and the LDM data sample was 154 times larger than the amount of training data from ADNI.

### C. Optimization of Model Hyperparameters

We conducted a literature search to define the overall bounds of the different hyperparameters typically used to train a vision transformer architecture [16] [8] [9] [19]. We trained our models for 8-25 epochs, typically fewer epochs when fine-tuning a pre-trained model. We selected hyperparameters based on a random search within an upper and lower bound, including: learning rate $\epsilon$ {uniform distribution between 0.00001 to 0.001}, weight decay $\epsilon$ {uniform distribution between 0.00001 to 0.001}, warm-up epochs $\epsilon$ {1, 5, 10, 15}, number-of attention heads $\epsilon$ {2, 4, 8, 12}, and number of encoder layers $\epsilon$ {3, 4, 6}. A reduced learning rate was usually used to fine-tune the pre-trained models. Our other key hyperparameters included a batch size of 16 and the ADAM optimizer [20]. We found the optimal combination of hyperparameter values to be a warm epoch of 1, learning rate of 0.0000581, ADAM optimizer with weight decay of 0.5889, 12 attention heads and 6 encoder layers.

For our experiments we tested specific advanced training techniques including *mixup,* for run-time data augmentation [21]. *Mixup* is a technique to create combinations of pairs of training images based on a value, $\lambda$, obtained from a beta distribution. Our implementation was based on the approach taken in Singla et al. [16]. We randomly selected batches during each epoch of training to undergo augmentation. The augmented image, along with the target labels of the two original images are used to calculate the loss weighted by $\lambda$, similar to eqn. (1):

$$\dot{x} = \lambda x_i + (1 - \lambda) x_j \quad (1)$$

where $x_i$ and are $x_j$ input MRI scans and $\lambda \epsilon [0,1]$

Further, we evaluated the use of layer-wise learning rate decay as suggested in [22] to maximize feature learning and re-use during the fine-tuning stage. We also tested a commonly used [15] [16] learning strategy of warming-up for a few epochs followed by cosine-annealing. In line with previous works, which involves training transformers for a higher number of epochs, we also experimented with training slightly longer for about 50 epochs.

### D. Model Evaluation

Following training, the models were tested on two independent datasets ADNI and OASIS3. The models were evaluated via the receiver-operator characteristic curve-area under the curve (ROC-AUC) as well as the accuracy, precision, F1-score, sensitivity and specificity calculated from the ROC-curve using a threshold obtained with Youden's Index [23].

## IV. RESULTS

The performance comparison of the models for AD classification is shown in Table I. The best performance for AD classification was achieved by NiT with an ROC-AUC of 0.8921 using ADNI and zero shot test ROC-AUC of 0.8107 on OASIS3. The NiT achieved an ROC-AUC of 0.968 and the ViT-B/16 achieved 0.987 on the UKBB dataset, this was also used in the pre-training phase. Table I also presents different types of model training strategies such as, fine-tuning models pre-trained on real and synthetic MRI data, the use of data augmentation, learning rate decay, layer-wise learning rate. Table II presents the ADNI and OASIS3 test ROC-AUC as a function of the proportion of ADNI training data used. We used data from 10% (220 scans) through 100% (2577 scans) of the training dataset from ADNI.

TABLE I. TEST PERFORMANCE USING VISION TRANSFORMERS, WITH DIFFERENT TRAINING STRATEGIES ON THE ADNI AND OASIS DATASETS

| Architecture | Training data% | Pre-training on UKB37K (Real data) | Pre-training on LDM100K (Diffusion-based synthetic data) | Training techniques | Classification of Dementia | |
|---|---|---|---|---|---|---|
| | | | | | ADNI, Test ROC-AUC | OASIS3, zero-shot Test ROC-AUC |
| NiT | 100% | - | - | More attention heads | **0.8921** | 0.8107 |
| NiT | 100% | - | - | Run-time data augmentation | 0.8291 | 0.7636 |
| NiT | 25% | ✓ | - | - | 0.8049 | 0.7841 |
| ViT-B/16 | 25% | ✓ | - | - | 0.8032 | 0.7519 |
| ViT-B/16 | 25% | - | ✓ | Layer-wise LR decay | 0.7628 | 0.7500 |
| ViT-B/16 | 25% | - | ✓ | - | 0.7501 | 0.7239 |
| ViT-B/16 | 25% | - | - | - | 0.7108 | 0.6916 |
| NiT | 25% | - | - | - | 0.6011 | 0.6709 |

Fig. 2 plots the trends in Table II as a data-model scaling curve comparing the performance of the NiT (blue) with the Random Forest classifier using radiomics features (orange).

## V. DISCUSSION

In this paper we showcased the capabilities of the ViT-like models on neuroimaging tasks with a range of difficulty, sex classification (easier) and AD classification (more difficult). We achieved our best model performance with an ROC-AUC of 0.8921 when training with the most data available (2577 scans) and when trained using additional attention heads in the transformer encoder. Further, we show the usefulness of the features learnt by the NiT via a zero-shot performance on an independent dataset i.e., OASIS3 without any additional finetuning of the model on the data. From our experiments, we noticed the ViT-like models much like CNNs do better on the test data when trained on increasing proportions of data compared to classical machine learning methods. The net performance gains with training the NiT on more data when compared with the random forest classifier is shown in Fig. 2. We noticed the NiT's sensitivity to model hyperparameters like number of attention heads, encoder layers, learning rate, weight decay and training strategies like run-time data augmentation, learning rate annealing with warmups in line with prior works. In this paper, we trained our models usually up to 25 epochs to simulate fast training with limited data. In general, training for longer (up to 50 epochs) improved performance.

Future work, in line with the current literature, will include, training the vision transformers for longer and with additional data to further boost performance.

TABLE II. DATA EFFICIENCY, AS MEASURED BY PERFORMANCE WITH DIFFERENT PROPORTIONS OF TRAINING DATA FROM THE ADNI DATASET

| Performance Metric\Training Data % | 10% | 25% | 50% | 75% | 100% | 100% (trained longer) | 100% (more attention heads) |
|---|---|---|---|---|---|---|---|
| ROC-AUC | 0.6011 | 0.7194 | 0.6999 | 0.7439 | 0.8170 | 0.8291 | 0.8921 |
| Accuracy | 0.6120 | 0.6751 | 0.6620 | 0.7030 | 0.7826 | 0.7974 | 0.8269 |
| F1-score | 0.4239 | 0.6243 | 0.6142 | 0.6409 | 0.7039 | 0.7508 | 0.8059 |
| Precision | 0.5769 | 0.6150 | 0.5974 | 0.6605 | 0.8378 | 0.7881 | 0.7711 |
| Sensitivity | 0.3353 | 0.6339 | 0.6320 | 0.6224 | 0.6069 | 0.7168 | 0.8439 |
| Specificity | 0.8171 | 0.7057 | 0.6843 | 0.7629 | 0.9129 | 0.8571 | 0.8143 |

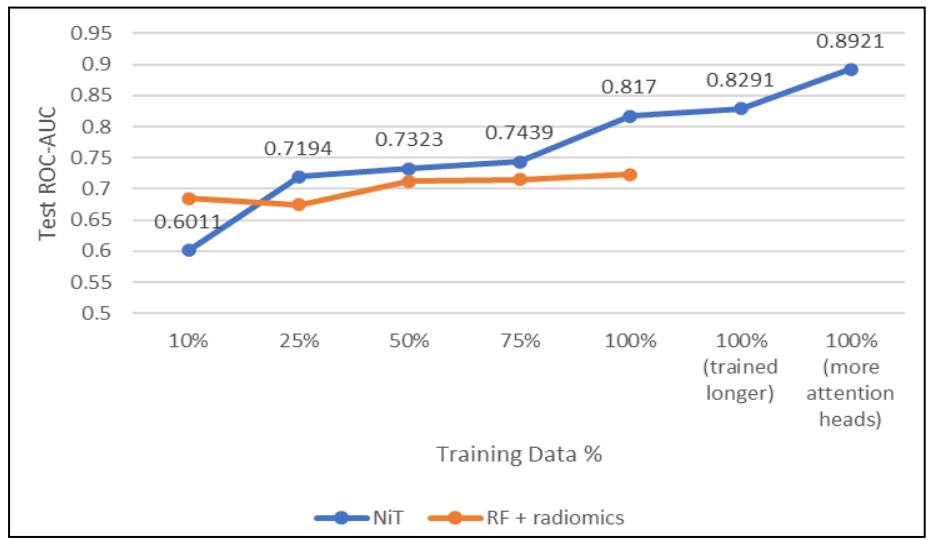

Figure 2. Data-model scaling curve with the NiT (blue). Curve shows the test performance for Alzheimer's disease classification on the ADNI dataset, as a function of the % of the training data that was used to train the NiT. Mix-up run-time data augmentation was used in most cases, and a learning rate with warm-up was used with the experiments using 50%, 75% of training data as part of the hyperparameter optimization. The NiT data-model scaling curve is compared with a random forest (RF) model trained on radiomics features (orange) extracted from the T1-w MRI scans [2]. With sufficient data, the transformer performs better. As expected, the trend line shows that the transformer performance is enhanced with additional training data with additional attention heads or with longer training times.

## VI. CONCLUSION

In this work, we evaluated the vision transformer class of architectures for Alzheimer's disease detection using T1-weighted MRI images. We present the effect of strategies like run-time augmentation, pretraining and other training techniques on the test-time performance in the context of working with limited data in the neuroimaging domain. We show using more training data in combination with the right set of model hyperparameters will allow researchers to use vision transformers for various challenging neuroimaging applications.

## ACKNOWLEDGMENT

This work was supported by the U.S. National Institutes of Health, under NIH grant U01 AG068057.